# Optical Constants of Photochemical Haze Analogs in $N_2$–$CH_4$–CO Atmospheres from 0.4 to 28.6 μm

Zhengbo Yang[1], Chao He[1,*], Haixin Li[1], Sai Wang[1], Xiao'ou Luo[1], Yu Liu[1,*], Sarah M. Hörst[2,3]

1. National Key Laboratory of Deep Space Exploration/School of Earth and Space Sciences, University of Science and Technology of China, Hefei 230026, China.

2. Department of Earth and Planetary Sciences, Johns Hopkins University, Baltimore, MD, USA.

3. Space Telescope Science Institute, Baltimore, MD, USA.

*Corresponding authors: Chao He (chaohe23@ustc.edu.cn), Yu Liu (yliu001@ustc.edu.cn)

## Abstract

Photochemical hazes play an important role in shaping the spectra and radiative balance of $N_2$-dominated planetary atmospheres. We present newly acquired FTIR and retrieved optical constants ($N = n + ik$) of laboratory-generated haze analogs from $N_2/CH_4$ and $N_2/CH_4/CO$ gas mixtures under plasma discharge conditions. The retrievals use particle densities newly measured for the $CH_4$-series and previously published particle densities for the CO-series. The experiments systematically explored $CH_4$ concentrations from 0.5% to 10% and CO concentrations from 0% to 5% with fixed 5% $CH_4$. Using measured particle densities together with the Beer–Lambert law and subtractive Kramers–Kronig (SKK) relation, we derived optical constants over the 350-25000 $cm^{-1}$ (0.4-28.6 μm) spectral range, with the 0.4–25 μm results presented in the main text. The infrared spectra reveal prominent absorption features associated with hydrocarbon-, nitrogen-, and oxygen-bearing functional groups. Increasing $CH_4$ abundance enhances aliphatic hydrocarbon features and corresponds to decreasing particle density, whereas increasing CO abundance promotes oxygen incorporation, broader mid-infrared absorptions, and higher particle density. The derived $k$ spectra exhibit strong absorptions near ∼3 μm, ∼4.6 μm, and ∼6-10 μm, while the real refractive index $n$ generally ranges from ∼1.2 to 1.7. The controlled $CH_4$- and CO-series establish composition-dependent variations in haze optical properties. A benchmark comparison among Titan-, Pluto-, and Triton-like haze analogs then uses these experimentally identified trends to interpret the optical differences among $N_2$-dominated planetary haze compositions. The density and optical constants provide laboratory constraints for atmospheric radiative transfer models and for interpreting planetary and exoplanetary spectra from spacecraft and telescopes.

## 1. Introduction

Photochemical hazes are widespread in reducing atmospheres and can affect planetary radiation budgets, thermal structures, and observed spectra. The most iconic example is Titan, Saturn's largest moon, where solar ultraviolet radiation and energetic particles dissociate $N_2$ and $CH_4$, initiating a cascade of complex organic chemistry that ultimately forms organic haze particles (D. Dubois 2025; S. M. Hörst 2017; C. A. Nixon 2024). Similar haze formation processes have also been identified or inferred for Pluto, Triton, and other volatile-rich bodies in the outer Solar System, where varying abundances of $CH_4$ and CO are expected to modify haze formation pathways and optical behavior (K. Arimatsu et al. 2026; M. E. Brown et al. 2011; P. Gao et al. 2017; G. R. Gladstone et al. 2016; J. I. Lunine 1989; S. E. Moran et al. 2022; S. Protopapa et al. 2025; S. C. Tegler et al. 2010). For instance, observations from New Horizons revealed that Pluto's hazes extend hundreds of kilometers above the surface and exhibit a blue color and strong forward scattering, consistent with fractal aggregates analogous to those on Titan, yet with distinct compositional differences due to the different composition and physical condition of Pluto's atmosphere (S. Fan et al. 2022; P. Gao, et al. 2017; G. R. Gladstone, et al. 2016; M. L. Wong et al. 2017). The relevance of photochemical hazes extends well beyond the Solar System. Observations and theoretical studies suggest that some exoplanets may host $N_2$-dominated atmospheres (e.g., R. Hu & H. D. Diaz 2019; C. V. Morley et al. 2017). In such environments, the relative abundance of $CH_4$ and CO—controlled by elemental composition, metallicity, and redox state—strongly influences atmospheric chemistry and may determine whether haze formation is favored (C. He et al. 2018; S. M. Hörst et al. 2018; S. Wang et al. 2026).

Atmospheric chemistry governs haze formation and composition, and therefore their optical properties. Haze optical properties affect the absorption and scattering of solar/stellar radiation, impacting the observation and assessment of planetary habitability (G. N. Arney et al. 2017; L. Corrales et al. 2023; C. He et al. 2024; P. Lavvas et al. 2010). Accurate treatment of such hazes requires knowledge of their optical constants, expressed as the complex refractive index ($n + ik$). These quantities are fundamental inputs for radiative-transfer calculations, climate models, and the interpretation of observational data. Haze optical constants vary across wavelengths, controlling atmospheric heating rates, spectral slopes, and continuum opacity (T. Drant et al. 2024; C. He, et al. 2024).

Despite their recognized importance, a systematic understanding of how carbon sources—specifically the $CH_4$/CO ratio—influence their optical constants remains limited. Due to the lack of direct constraints of haze properties in planetary atmospheres, haze analogs, called "tholins", are produced in laboratory experiments that simulate relevant planetary environments, and are investigated to constrain their chemical and physical properties (M. L. Cable et al. 2012; C. Sagan & B. N. Khare 1979). A number of laboratory studies (**Table 1**) have reported optical constants for tholin samples produced under various conditions, such as different energy sources, pressures, temperatures, and gas compositions (e.g., C. He et al. 2022a; H. Imanaka et al. 2012; B. N. Khare et al. 1984; A. Mahjoub et al. 2012; E. Sciamma-O'Brien et al. 2012). While these studies provide valuable reference data, the diversity of experimental setups/conditions and measurement approaches makes direct comparison difficult. Existing datasets were derived using different retrieval methods and often cover different wavelength ranges. At the same time, the abundances of $CH_4$ and/or CO vary substantially among different planetary atmospheres and can also change with altitude within a single atmosphere, potentially altering atmospheric chemistry and the composition and optical properties of the resulting hazes. Previous laboratory studies have shown that $CH_4$ abundance can affect haze composition and optical behavior, while CO-containing or more oxidized gas mixtures promote oxygen-bearing haze chemistry and composition-dependent optical properties (T. Drant, et al. 2024; L. Jovanović, et al. 2021; A. Mahjoub, et al. 2012; S. E. Moran, et al. 2022). A recent cross-laboratory study further showed that both the initial gas composition and experimental conditions can affect the retrieved refractive indices (T. Drant, et al. 2026). C. He, et al. (2022a) derived optical constant for a $N_2$/$CH_4$ haze analog via vacuum FTIR spectroscopy, minimizing the spectral interference from Earth's atmosphere. However, published studies differ in their experimental conditions, sample forms, wavelength coverage, optical methods, and the composition parameters varied. The present work provides a directly comparable broadband dataset for controlled $CH_4$- and CO-series produced in the same experimental framework.

**Table 1.** Overview of previous laboratory measurements of optical constants for photochemical haze analogs, including experimental conditions, gas compositions, spectral coverage, as well as measurement and derivation methods.

| Study | Setup | Gas composition (%) | Experimental conditions | | | Wavelength range | Measurement and derivation method |
|---|---|---|---|---|---|---|---|
| | | | Energy source | Pressure | Temperature | | |
| B. N. Khare, et al. (1984) | Ionization chamber (Cornell University) | $N_2/CH_4$ (90/10) | DC discharge 0.2 kV/15 mA | 0.2 mbar (3 sccm) | Room temperature | 0.025-1000 μm | Spectrometry Ellipsometry Kramers-Kronig analyses |
| S. Ramirez (2002) | LISA chamber (Laboratoire Interuniversitaire des Systemes Atmosph ` eriques) | $N_2/CH_4$ (98/2) | DC discharge 4 kV/82-92 mA | 2 mbar (3 sccm) | Room temperature | 0.2-0.9 μm | Spectrometry Abelès equations |
| B. N. Tran et al. (2003) | The photochemical flow reactor (Rensselaer Polytechnic Institute) | $N_2/CH_4/H_2$ (98/1.8/0.2) + $C_2H_2/C_2H_4/HC_3N$ (350/300/17 ppm)<br>$N_2/CH_4/H_2$ (98/1.8/0.2) + $C_2H_2/C_2H_4/HC_3N$ (35/30/1.7 ppm)<br>$N_2/CH_4/H_2$ (98/1.8/0.2) + $C_2H_2/C_2H_4/HC_3N$ (350/300/17 ppm)<br>$N_2/C_2H_4$ (3 ppm) | UV source: external low pressure mercury lamp (185 and 254 nm) | 933 mbar (15 sccm)<br>933 mbar (15 sccm)<br>133 mbar (15 sccm)<br>933 mbar (15 sccm) | Room temperature | 0.2-2.5 μm<br>2.5-15 μm | Spectrometry Ellipsometry Iteration approach for fitting measured T and R |
| V. Vuitton et al. (2009) | The photochemical flow reactor (Rensselaer Polytechnic Institute) | $N_2/CH_4/H_2$ (98/1.8/0.2) + $C_2H_2/C_2H_4/HC_3N$ (350/300/17 ppm)<br>$N_2/CH_4/H_2$ (98/1.8/0.2) + $C_2H_2/C_2H_4/HC_3N$ (3500/3000/170 ppm)<br>$N_2/CH_4/H_2$ (98/1.8/0.2) + $C_2H_2/C_2H_4/HC_3N/CO$ (350/300/17/3000 ppm) | UV source: external low pressure mercury lamp (185 and 254 nm) | 933 mbar (15 sccm) | Room temperature | 0.375-1.55 μm | Photothermal deflection spectroscopy Matrix approach outlined in Klein and Furtak (1986) |
| C. A. Hasenkopf et al. (2010) | Flow system (CU Boulder) | $N_2/CH_4$ (99.9/0.1) | deuterium continuum UV lamp | 840 mbar (60 sccm) | Room temperature | 0.532 μm | Cavity-ringdown aerosol extinction spectroscopy (CRD-AES) Mie theory |
| H. Imanaka, et al. (2012) | ICP reactor (NASA Ames Research Center) | $N_2/CH_4$ (90/10) | RF plasma 100W | 0.26-1.6 mbar | Room temperature | 1.47-25 μm | Spectrometry Iterative subtractive Kramers–Kronig (SKK) |
| E. Sciamma-O'Brien, et al. (2012) | PAMPRE chamber (LATMOS) | $N_2/CH_4$ (95/5) | RF plasma 30 W | 0.9 mbar (55 sccm) | Room temperature | 0.37-0.9 μm | Spectroscopic ellipsometry Kramers–Kronig Relation |
| A. Mahjoub, et al. (2012) | PAMPRE chamber (LATMOS) | $N_2/CH_4$ (99/1)<br>$N_2/CH_4$ (98/2)<br>$N_2/CH_4$ (95/5)<br>$N_2/CH_4$ (90/10) | RF plasma 30 W | 0.9 mbar (55 sccm) | Room temperature | 0.37-1 μm | Spectroscopic ellipsometry Tauc–Lorentz model |
| L. Gavilan et al. (2018) | PAMPRE chamber (LATMOS) | $N_2/CH_4$ (95/5)<br>$N_2/CH_4/CO_2$ (90/5/5)<br>$N_2/CH_4/CO_2$ (90/4/6)<br>$N_2/CH_4/CO_2$ (95/2.5/2.5)<br>$N_2/CH_4/CO_2$ (95/2/3)<br>$N_2/CH_4/CO_2$ (95/2/8)<br>$N_2/CH_4/CO_2$ (95/1/4) | RF plasma 30 W | 0.95 mbar | ~340 K | 0.13-10 μm | Spectrometry Only *k* derived from Lambert's law |
| L. Jovanović et al. (2021) | PAMPRE chamber (LATMOS) | $N_2/CH_4/CO$ (99.5/0.5/500 ppm)<br>$N_2/CH_4/CO$ (99/1/500 ppm)<br>$N_2/CH_4/CO$ (95/5/500 ppm) | RF plasma 30 W | 0.9±0.1 mbar (55 sccm) | Room temperature | 0.27-2.1 μm | Spectroscopic ellipsometry Tauc–Lorentz dispersion model |
| C. He, et al. (2022a) | PHAZER chamber (Johns Hopkins University) | $N_2/CH_4$ (95/5) | AC glow discharge | 2.66 mbar (10 sccm) | ~100 K | 0.4-3.5 μm | Spectrometry subtractive Kramers–Kronig (SKK) |
| E. Sciamma-O'Brien et al. (2023) | COSmIC chamber (NASA Ames Research Center) | $N_2/CH_4$ (95/5)<br>$N_2/CH_4/C_2H_2$ (94.5/5/0.5)<br>$Ar/CH_4$ (95/5) | pulsed DC plasma discharge (10 Hz, 700-1000 V) | <30 mbar (2000 sccm) | <150 K | 0.4-1.6 μm | Spectrometry Modified Cauchy dispersion model |
| L. Corrales, et al. (2023) | PAMPRE chamber (LATMOS) | $N_2/CH_4$ (95/5)<br>$N_2/CH_4/CO_2$ (90/5/5)<br>$N_2/CH_4/CO_2$ (95/2/8) | RF plasma 30 W | 0.95 mbar | ~340 K | 0.13-10 μm | Spectrometry Kramers–Kronig relations |
| T. Drant, et al. (2024) | PAMPRE chamber (LATMOS) | $N_2/CH_4/CO_2$ (95/4/1)<br>$N_2/CH_4/CO_2$ (95/2/3) | RF plasma 30 W | 0.85 mbar (60 sccm) | Room temperature | 0.3-30 μm | Spectrophotometry ellipsometry Kramers-Kronig equation |
| T. Drant et al. (2026) | PAMPRE chamber (LATMOS) and COSmIC chamber (NASA Ames Research Center) | $N_2/CH_4$ (95/5)<br>$N_2/CH_4$ (90/10)<br>$N_2/CH_4/CO$ (95/4.95/500 ppm)<br>$N_2/CH_4/CO$ (95/4/1)<br>$Ar/CH_4$ (95/5)<br>$Ar/CH_4/CO$ (95/4/1) | RF plasma, 30 W (PAMPRE); pulsed DC plasma (COSmIC) | ~1 mbar (PAMPRE); ~30 mbar (COSmIC) | ~300 K (PAMPRE); ~200 K (COSmIC) | 0.25-200 μm | Spectroscopy, ellipsometry, and FTIR spectroscopy Iterative fitting subtractive Kramers–Kronig (SKK) |

To address this gap, we investigate laboratory-generated haze analogs produced under fixed temperature and pressure conditions while varying only $CH_4$ and CO ratios in $N_2$-dominated gas mixtures. Using vacuum Fourier-transform infrared (FTIR) spectroscopy, we measure the transmittance spectra of the resulting haze analogs over a broad wavelength range (0.4–28.6 μm). The optical constants ($n$, $k$) are then retrieved through iterative fitting and subtractive Kramers–Kronig analysis, following methods established in recent work (C. He, et al. 2024; H. Li et al. 2025). By holding all other experimental parameters constant except for the $CH_4$ or CO ratios, any measured difference in optical properties can be directly linked to changes in the gas-phase chemistry and the composition of the resulting solids. These compositional variations in turn modify haze absorption and scattering, with important consequences for atmospheric radiative transfer. By providing a self-consistent set of optical constants for haze analogs formed under varying carbon-source conditions, this work establishes a critical foundation for improving radiative-transfer models and spectral retrievals for diverse planetary atmospheres, from Titan, Triton, and Pluto to many exoplanets.

## 2. Materials and Methods

### 2.1 Haze Analog Production

The CO and $CH_4$ compositional-series haze analogs in this study were produced under controlled temperature and pressure conditions using the Planetary HAZE Research (PHAZER) experimental setup (**Figure 1**), which is a stainless-steel chamber equipped with a glow discharge system, capable of simulating chemistry of a range of planetary atmospheres. A detailed description of the setup and operating procedures can be found in previous studies (C. He, et al. 2018; C. He et al. 2017).

To investigate the effects of varying carbon sources on photochemical haze formation and properties, we designed two series of experiments. The first series examined the influence of methane abundance without CO, using $N_2$–$CH_4$ gas mixtures with $CH_4$ concentrations of 0.5%, 1%, 2%, 5%, and 10% - we will thereafter call it $CH_4$-series (**Figure 1**). The second series investigated the role of CO by fixing $CH_4$ at 5% and varying CO concentrations at 0%, 0.05%, 0.2%, 0.5%, 1%, 2.5%, and 5%, balanced with $N_2$ - thereafter called CO-series (**Figure 1**). These gas mixtures were prepared using ultra-high-purity gases ($N_2$: 99.9995%, $CH_4$: 99.999%, CO: 99.99%, Airgas) in a dedicated mixing manifold. The gas mixture flowed through a 15-meter cooling coil and was cooled to ~100 K before being introduced into the reaction chamber. The flow rate was 10 standard cubic centimeters per minute (sccm), maintaining the pressure in the reaction chamber at approximately 2 Torr throughout the experiments.

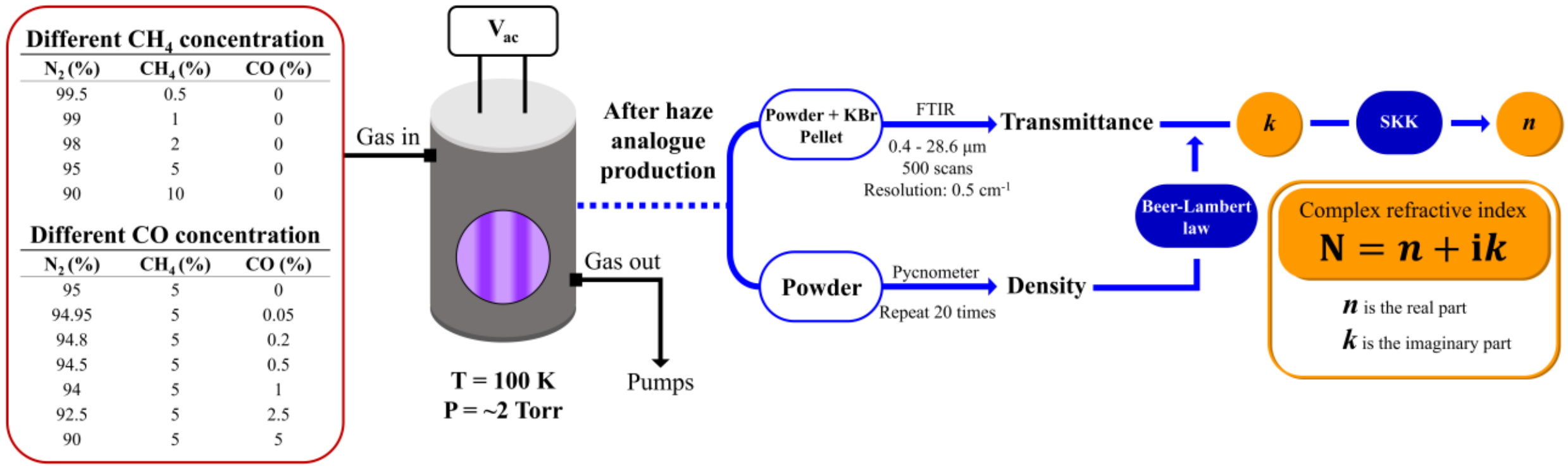

**Figure 1.** Simplified schematic of the experimental setup and the analytical procedure for this study. Gas mixtures with varying $CH_4$ or CO concentrations are introduced into the PHAZER chamber and processed under AC plasma conditions to produce haze analog particles. The resulting solid products are collected as powders and subsequently analyzed using two approaches. For optical measurements, powders are mixed with KBr to form pellets and measured by FTIR spectroscopy over 0.4-28.6 μm to obtain transmittance spectra. Particle densities are measured using a pycnometer. Insets list the gas compositions for the $CH_4$-and CO- series explored in this work. For the present study, broadband FTIR spectra and $CH_4$-series particle densities were newly measured from archived powders; CO-series particle densities were adopted from C. He, et al. (2017).

A glow discharge was ignited by applying a 15000 V alternating current (AC) power supply to the electrodes, initiating the dissociation and ionization of precursor gases and subsequent formation of gas-phase products and solid organic haze particles. The resulting solid products were deposited on the chamber walls. After each experiment, the chamber was evacuated to $<10^{-4}$ Torr before sample collection. All samples were handled in a nitrogen-filled glovebox ($O_2$, $H_2O$ < 0.1 ppm) to prevent atmospheric contamination.

No haze samples were newly produced for the present study. The CO and $CH_4$ compositional-series haze analogs used in this study were produced in 2016. The Triton-like benchmark sample was produced separately in 2022. Following collection in the glovebox, the samples were stored in sealed sample vials secured with Parafilm, wrapped in aluminum foil to exclude light, and kept in an $N_2$ glovebox to minimize exposure to $O_2$ and $H_2O$ prior to the optical property measurement reported here. Previous characterizations were performed on aliquot material from these same production batches. The CO-series samples were previously characterized for haze production rate, particle density, elemental composition, and molecular structure by NMR spectroscopy (C. He, et al. 2017). For the $CH_4$-series samples, we obtained the haze production rate and elemental composition (C. He et al. 2022b). Particularly, we also measured the transmittance and reflectance from 0.4 to 3.5 μm of the 5% $CH_4$/95% $N_2$ sample using haze film on quartz substrate (C. He, et al. 2022a). For the Triton-like sample, S. E. Moran, et al. (2022) reported the haze production rate, elemental composition, molecular composition based on high-resolution mass spectrometry, and transmittance and reflectance of haze films on quartz substrates from 0.4 to 5 μm. These previous measurements provide the chemical and physical context for interpreting the optical-property trends examined here. In this study, we newly measured the particle densities of the $CH_4$-series samples and Triton-like sample; for all the haze analogs, we acquired the broadband (0.4-28.5 μm) FTIR spectra using the KBr pellet method. The measurement procedure is described below.

## 2.2 Haze Analog Characterization

The archived powders from each compositional condition were used for the present density and spectra measurements. The densities for the $CH_4$-series were newly measured using a gas pycnometer (AccuPyc II 1340, Micromeritics) as described in previous studies (C. He, et al. 2017; H. Li, et al. 2025). In contrast, the particle densities for the CO-series were adopted from C. He, et al. (2017). The volume was determined from helium gas displacement using the ideal gas law. A precisely weighed amount of haze powder (typically ~50 mg) was placed in a sealed calibrated cell and measured 20 times to obtain an average density and standard deviation**.**

For the transmittance measurements, the collected haze powders were prepared as haze/KBr pellets. All handling was performed inside the $N_2$ glovebox to avoid moisture uptake. KBr was dried at 110 °C for 12 h prior to use. Haze powder was mixed with KBr at a target mass fraction of ~0.3 wt% (haze/KBr). This target concentration was selected to balance sufficient absorption signal against minimizing particle-induced scattering. To minimize gravimetric errors, we use a two-step dilution procedure: first a ~1.5 wt% master mixture was prepared by ball-milling, and a portion of this mixture was then diluted further with KBr to the target concentration. The actual haze mass for each individual pellet, rather than the nominal target mass fraction, was used in the optical-constant retrieval. This dilute-pellet strategy is consistent with recent KBr-pellet optical-constant measurements of organic solids (K. A. Peterson et al. 2024). Approximately 200 mg of the final mixture was pressed in a 13 mm diameter stainless steel die using a hydraulic press with a staged force protocol: 3 tons for 2 s, 7 tons for 30 s, and finally 10 tons for 120 s. The resulting pellets were ~0.5 mm thick and free of cracks. A freshly prepared pure KBr pellet was measured using the identical configuration on the same day as the corresponding haze-pellet measurements to minimize the effects of instrument drift and time-dependent background variation. The measured transmittance of the pure KBr pellets ranges from 0.85 to 0.91 across the measured wavelength range (0.4-28.6 μm), and these spectra were used as the reference for the corresponding haze-pellet transmittance measurements.

Broadband transmittance spectra of the archived haze samples were newly acquired using a Bruker Vertex 70v Fourier-transform infrared spectrometer (FTIR) operating under vacuum (pressure $< 0.2$ mbar) to minimize absorption from atmospheric $H_2O$ and $CO_2$. The spectrometer was configured with interchangeable beamsplitters (quartz and KBr) and detectors (silicon diode for the visible–near-infrared region and DLaTGS for the mid-infrared), enabling continuous coverage from 0.4-28.5 μm (25,000 to 350 $cm^{-1}$). All measurements were performed at room temperature (294 K). For each sample, 500 scans were acquired at a spectral resolution of 0.5 $cm^{-1}$ to improve signal-to-noise ratio. The transmittance spectrum of each sample was obtained by ratioing the sample intensity against the KBr reference spectrum. At least three measurements were taken from different pellet regions to ensure reproducibility, and the averaged spectra were used for subsequent analysis. Although the FTIR measurements extend to 350 $cm^{-1}$ (28.5 μm), the spectra near the detector edge below ~400 $cm^{-1}$ are associated with larger instrumental uncertainties and are therefore less reliable. Therefore, we only present all spectra and derived optical constants in the range of 25000–400 $cm^{-1}$ (0.4–25 μm). The full dataset down to 350 $cm^{-1}$ is provided in the supplementary materials.

### 2.3 Derivation Method of Optical Constants

The complex refractive index ($N = n + ik$) of the haze analogs was derived from the measured transmittance spectra using a combination of the Beer–Lambert law and the subtractive Kramers–Kronig (SKK) relation, following the methodology established in recent work (C. He, et al. 2024; H. Li, et al. 2025). As discussed below, applying the Beer–Lambert law to haze-containing KBr pellets is an approximation, and its limitations are considered when interpreting the absolute optical constants.

The effective thickness ($d$) of the haze material in the KBr pellet was calculated as:

$$d = \frac{m}{\pi r^2 \rho}$$

where $m$ is the recorded mass of haze particles in the individual pellet, $r$ is the pellet radius (6.5 mm), and $\rho$ is the measured density of the haze sample. The extinction coefficient $k(\nu)$ was then obtained via the Beer–Lambert law:

$$k(\nu) = \frac{1}{4\pi\nu d} \ln(\frac{1}{T(\nu)})$$

where ν is the wavenumber in $cm^{-1}$ and $T(\nu)$ is the measured transmittance. For the Beer-Lambert law approximation to be applicable, particle scattering and other residual non-absorptive losses, including reflections at pellet interfaces, should be minimized. A pure KBr pellet reference accounts for instrumental response and contributions associated with the KBr matrix and pellet preparation, but it does not fully remove scattering introduced by the haze particles themselves. We therefore prepared the haze/KBr pellets at a sufficiently dilute concentration to minimize these effects, consistent with previous discussions of the assumptions and limitations of KBr-pellet measurements (T. G. Mayerhofer et al. 2016; T. G. Mayerhofer et al. 2020; K. A. Peterson, et al. 2024). For the retrieved imaginary refractive index $k$, we propagated uncertainties associated with the FTIR transmission measurement, haze mass used in the two-step dilution and pellet preparation, and haze particle density. The uncertainty in the transmission ratio includes the relative uncertainties of both the haze/KBr and pure-KBr reference spectra. The resulting wavelength-dependent propagated uncertainty in k is shown as a shaded ±1σ band in the optical-constant figures and is included in the data products. Potential systematic effects associated with residual particle scattering and departures from the Beer–Lambert approximation are discussed separately above.

The real part of the refractive index $n(\nu)$ was derived using the subtractive Kramers–Kronig (SKK) relation (H. Imanaka, et al. 2012; G. D. McDonald et al. 1994; B. E. Wood & J. A. Roux 1982):

$$n(\nu) = n_0 + \frac{2(\nu^2 - \nu_0^2)}{\pi} \mathrm{P} \int_0^\infty \frac{\nu' k(\nu')}{(\nu'^2 - \nu^2)(\nu'^2 - \nu_0^2)} d\nu'$$

where P denotes the Cauchy principal value, $\nu_0$ is a reference wavenumber and $n_0 = n(\nu_0)$ is the anchor value. We choose the 15000–17000 $cm^{-1}$ interval as the candidate anchoring region because it lies within the visible range, where the optical response is comparatively smooth and away from the strong absorption features in the measured spectra. More importantly, an independently determined refractive-index spectrum is available in this spectral region from C. He, et al. (2022a), providing an external reference for the SKK retrieval. To determine whether any individual frequency within this interval was preferable as the anchor, we examined the wavelength-dependent uncertainty of $n$ reported by C. He, et al. (2022a). The reported uncertainties (4~5%) are comparable across this frequency range and no individual frequency is clearly better constrained than the others. Therefore, we select $\nu_0$ = 16000 $cm^{-1}$as the final anchor frequency, as it lies near the center of the justified interval. This selection is independent of the anchor-frequency sensitivity test described below.

To evaluate the sensitivity of the SKK retrieval to the choice of a reasonable anchor frequency, we re-performed the retrieval using 51 reference frequencies from 15000 to 17000 $cm^{-1}$, spaced by 40 $cm^{-1}$. For each trial frequency, we adopted the $n$ value of the nearest frequency reported by C. He et al. (2022a) as the corresponding anchor value. In this test, the maximum difference between the 16th- and 84th-percentile retrievals is less than 0.003 over the reported wavelengths, showing that the wavelength-dependent $n$ spectra are only weakly sensitive to the choice of anchor frequency within this interval. This sensitivity test is a robustness diagnostic and is not treated as an independent experimental uncertainty of the reported $n$ values. For all samples, we adopted $n_0$ = 1.56 at $\nu_0$ = 16000 $cm^{-1}$ obtained by the wavelength-dependent $n$ spectrum reported by C. He, et al. (2022a), where the $n$ value was determined using the Kramers–Kronig relation for the 5% $CH_4$/95% $N_2$ haze film sample. This value is consistent with the range of $n$ values reported for similar haze analogs (A. Mahjoub, et al. 2012). The retrieved $n$ values are constrained by the uncertainty of the independently determined anchor value $n_0$, which is approximately 4.5% (0.069) at 16000 $cm^{-1}$ as reported in C. He, et al. (2022a). Because the same anchor value is adopted for all samples, this anchor uncertainty is common to all retrieved spectra rather than being sample-specific. Within the SKK retrieval, the contribution of $k$ uncertainty propagated to $n$ is negligible (<0.002) compared with the uncertainty associated with the anchor value. The common-anchor approach therefore enables internally consistent comparisons of the wavelength-dependent dispersion of each samples, but limits direct comparison of the retrieved n spectra among different samples because the true refractive index at the anchor frequency may differ between samples. The derivation procedure was applied to all the haze analogs in this study. This systematic series of experiments, covering both $CH_4$-dominated and CO-influenced conditions, allows us to isolate the effects of methane and carbon monoxide on the optical properties of haze analogs relevant to $N_2$-dominated reducing atmospheres, providing a self-consistent dataset for atmospheric modeling and observational interpretation.

## 3. Results and Discussion

### 3.1 Density of Haze Analogs

The newly measured $CH_4$-series particle densities and the CO-series particle-density values adopted from C. He, et al. (2017), are shown in **Figure 2**. The reported uncertainties are small (±0.0015-0.0029 $g/cm^3$) relative to the observed variations across gas mixtures, indicating that the observed differences are substantially larger than the measurement repeatability. Across all samples, densities span a relatively narrow range of 1.338–1.424 $g/cm^3$. The change in density reflects the compositional difference of the haze samples. For the $CH_4$-series samples (0% CO), the density decreases monotonically with increasing $CH_4$ fraction in the initial gas mixture, from 1.415 $g/cm^3$ at 0.5% $CH_4$ to 1.338 $g/cm^3$ at 10% $CH_4$. The variation is systematic across the full $CH_4$ range, with a total decrease of ~5%. This trend could reveal the progressive shift in chemical and physical properties of the haze particles. Previous laboratory studies have shown that methane concentration can significantly influence both the chemical composition and physical structure of haze particles (M. L. Cable, et al. 2012; D. Dubois et al. 2020; E. Sciamma-O'Brien et al. 2010). In particular, S. M. Hörst and M. A. Tolbert (2013) suggested that density variations in Titan haze analogs may reflect differences in particle porosity and aggregate structure in addition to bulk composition. For our $CH_4$-series experiments, we previously reported the haze production rates and their elemental compositions (C. He, et al. 2022b). We found that the N/C ratio decreases systematically with increasing $CH_4$ concentration, from ~0.97 at 0.5% $CH_4$ to ~0.66 at 10% $CH_4$, indicating reduced nitrogen incorporation and enhanced carbon-rich chemistry at higher methane abundance.

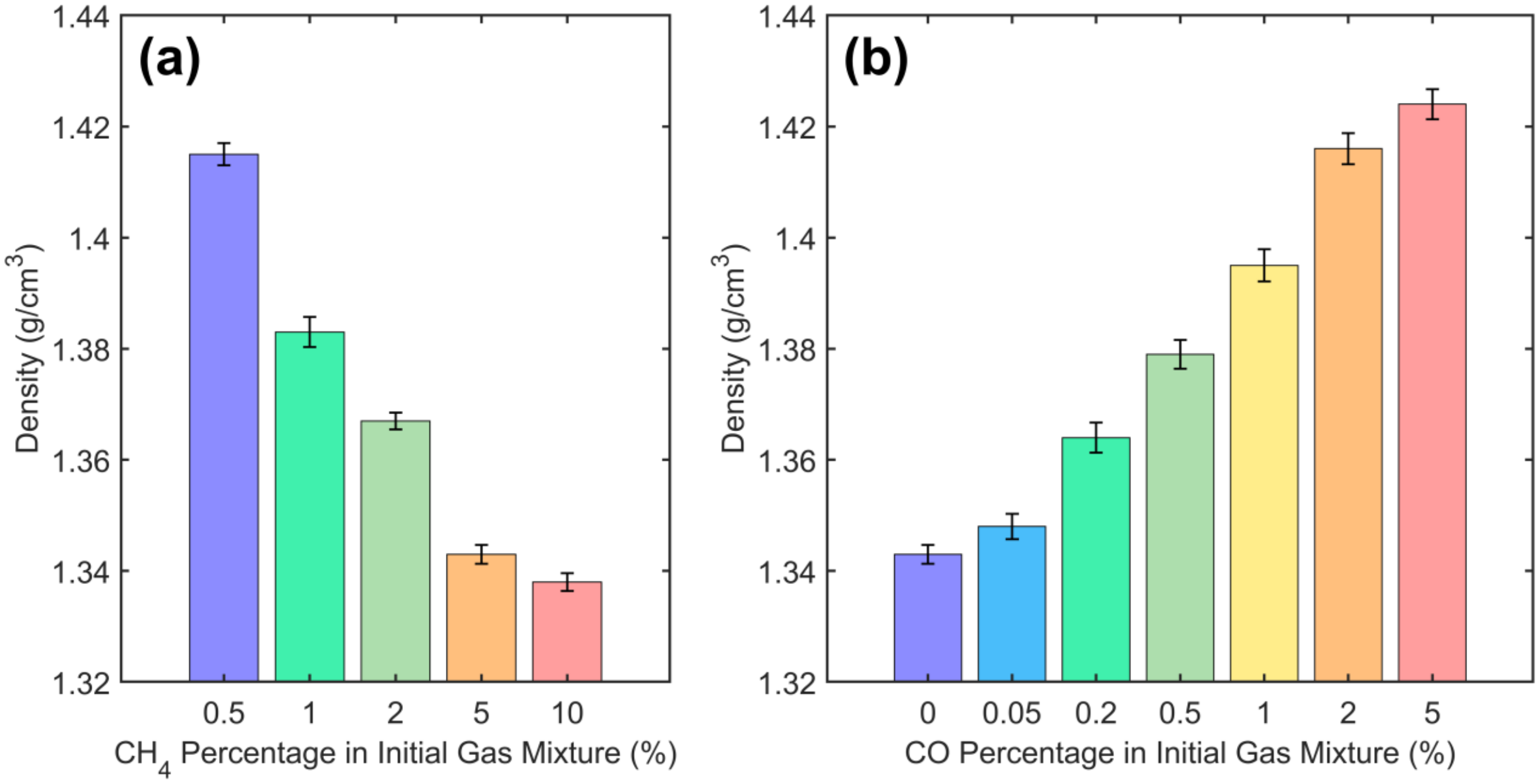

**Figure 2.** Measured densities of haze analog particles produced under different initial gas compositions. The left panel shows the variation in particle density as a function of increasing $CH_4$ concentration in the initial $N_2/CH_4$ gas mixtures. The right panel shows the variation in solid density with increasing CO concentration in the initial $N_2/CH_4/CO$ gas mixtures. Error bars represent the standard deviations derived from repeated pycnometer measurements. $CH_4$-series particle densities were newly measured in this work; CO-series particle densities were adopted from C. He, et al. (2017) and were used as input parameters in the present optical-constant retrieval. The full density dataset is available in the "Data behind Figure" files.

Since nitrogen-bearing functional groups can contribute to intermolecular interactions and hydrogen-bonding networks within the solid material, reduced nitrogen incorporation may decrease the compactness of the resulting organic solids. At the same time, higher $CH_4$ concentrations are associated with increased haze production rates, which may favor the formation of less compact materials. These combined effects likely contribute to the observed decrease in particle density. The compositional evolution toward more aliphatic-rich material is further supported by the FTIR spectral trends discussed in **Section 3.2**.

For the CO-series, we adopted the density values previously reported by C. He, et al. (2017) for the same original production batches used for the present FTIR measurements. The haze density increases with CO concentration (0%-5%), from 1.343 g/cm$^3$ at 0% CO to 1.424 g/cm$^3$ at 5% CO, while the haze production rate remains relatively stable. This trend suggests that the density enhancement is driven primarily by compositional and structural evolution rather than changes in particle yield. Elemental analyses showed that increasing CO abundance promotes oxygen incorporation into the haze particles, resulting in higher O/C ratios and more oxygen-rich organic material. The presence of oxygen-bearing functional groups likely increases the polarity and intermolecular interactions within the solids, favoring more compact molecular packing and therefore higher particle densities. However, C. He, et al. (2017) also noted that oxygen enrichment alone cannot fully explain the observed density increase, implying that additional structural factors, such as enhanced unsaturation and polarity, may also contribute.

The particle-density data provide direct constraints for the material properties of haze analog particles formed under $N_2/CH_4/CO$ conditions. The systematic dependence on gas composition implies that particle density should not be treated as a fixed parameter when simulating atmospheres with varying $CH_4$ and CO abundances. These density measurements can be incorporated as input parameters in microphysical and aerosol evolution models, where particle density directly affects sedimentation velocities, coagulation rates, and particle growth timescales. In particular, the observed decrease in density with increasing $CH_4$ and increase with CO suggest that compositional variations may lead to measurable differences in particle transport and vertical distribution. The density range derived here (1.34-1.42 $g/cm^3$) provides a constrained interval for use in such models, reducing uncertainties associated with assumed material properties.

### 3.2 FTIR Spectra of Haze Analogs

The FTIR spectra of solid products generated from $N_2/CH_4$ gas mixtures with varying $CH_4$ concentrations (0.5%-10%) are shown in **Figure 3a**. The full spectral range (0.4-25 μm, or 25000-400 $cm^{-1}$) exhibits a generally increasing absorption toward longer wavelengths, while the zoomed-in region (4000-400 $cm^{-1}$) reveals distinct vibrational features associated with various functional groups. Functional-group identifications were guided by the characteristic infrared frequencies summarized in D. Lin-Vien et al. (1991).

Several major absorption regions are consistently observed across all samples. A broad absorption feature spanning 3500-2500 $cm^{-1}$ is observed in all $CH_4$-series samples, with two prominent maxima near ~3300 $cm^{-1}$, characteristic of N-H stretching vibrations (3350-3280 and 3215-3190 $cm^{-1}$). Although O-H stretching (3500-3200 $cm^{-1}$ in alcohols and 3300-2500 $cm^{-1}$ in carboxylic acids) may also contribute within this spectral region, elemental analyses performed on the same samples by C. He, et al. (2022b) showed that the oxygen abundance remains below 3% in the haze analogs across the full $CH_4$ range, indicating that oxygen-bearing functional groups likely contribute only weakly to the observed spectra. The double-peaked structure also suggests that N-H stretching is the dominant contributor in this region. In the C-H stretching region (3000-2800 $cm^{-1}$), distinct peaks corresponding to $-CH_3$ antisymmetric (~2972 $cm^{-1}$), $-CH_2$ antisymmetric (~2932 $cm^{-1}$), and $-CH_3$ symmetric (~2872 $cm^{-1}$) stretching vibrations are clearly visible. Absorptions in the 2250-2150 $cm^{-1}$ region probably indicate the presence of triple-bonded species such as -C≡N (2250-2230 $cm^{-1}$), -N≡C (2146-2134 $cm^{-1}$), and C≡C (2310–2150 $cm^{-1}$), and cumulated double bonds like -N=C=N- (2152-2128 $cm^{-1}$) and C=C=N (~2018 $cm^{-1}$). The region between 1800 and 1500 $cm^{-1}$ shows multiple overlapping absorptions associated with double bonds and nitrogen-bearing functional groups. Prominent contributions include C=C stretching (~1650 $cm^{-1}$), C=N stretching (1640-1590 $cm^{-1}$), N=N (1576-1565 $cm^{-1}$), and $-NH_2$ scissoring mode (1627-1590 $cm^{-1}$). Although, oxygen-containing species such as C=O stretching (1800-1550 $cm^{-1}$), N=O (1621-1539 $cm^{-1}$), and $-NO_2$ (1601–1531 $cm^{-1}$) may also contribute within this region, their relative contribution is expected to be limited by the low oxygen abundance of the samples. In the fingerprint region (1500-1400 $cm^{-1}$), bands near 1475-1445 $cm^{-1}$ and 1395-1365 $cm^{-1}$ correspond to $-CH_2$ scissoring and $-CH_3$ bending modes, respectively. The absorption band observed in the region of 1420-1400 $cm^{-1}$ is assigned to the scissoring vibration of the =C-H bond. The bands between 1250 and 1000 $cm^{-1}$ are attributed to C-N stretching, with possible minor contributions from C-O stretching modes (1150-1060 $cm^{-1}$ for aliphatic ethers, 1225-1200 $cm^{-1}$ for vinyl ethers, and 1310-1210 $cm^{-1}$ for aromatic ethers) are also present. A weak feature near ~1111 $cm^{-1}$ is consistent with N-N stretching vibrations. In the lower wavenumber region, features near 681-610 $cm^{-1}$ can be assigned to ≡C-H bending modes.

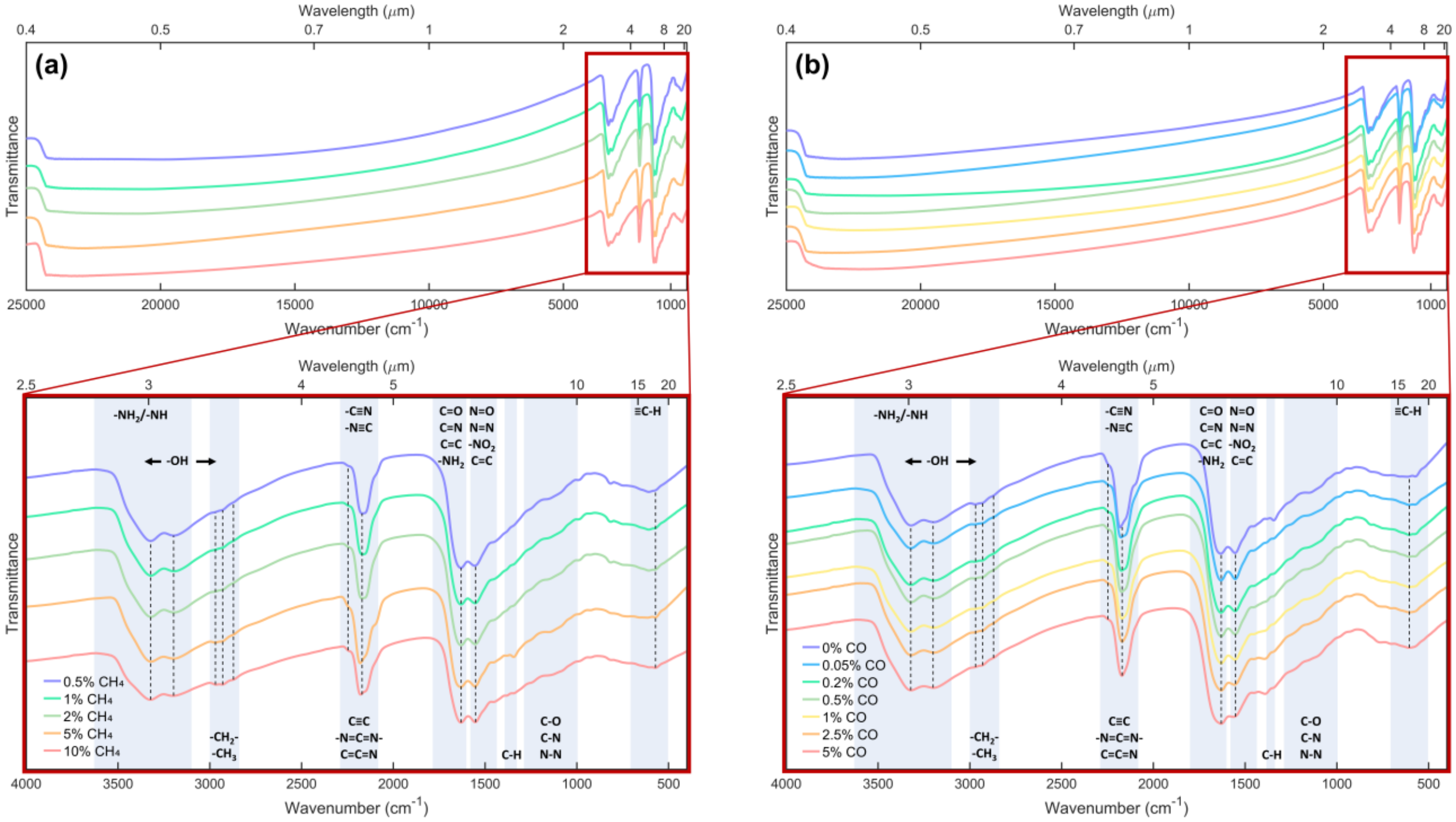

**Figure 3.** FTIR spectra of haze analog particles produced under different initial gas compositions. (a) FTIR spectra of solid products formed from $N_2/CH_4$ gas mixtures with $CH_4$ concentrations of 0.5%, 1%, 2%, 5%, and 10%. The upper panel shows the full spectral range from 25000 to 400 $cm^{-1}$ (0.4-25 μm), while the lower panel presents an enlarged view of the 4000-400 $cm^{-1}$ region. (b) FTIR spectra of solid products formed from $N_2/CH_4/CO$ gas mixtures with fixed $CH_4$ concentration (5%) and varying CO concentrations of 0%, 0.05%, 0.2%, 0.5%, 1%, 2.5%, and 5%. The upper panel shows the full spectral range from 25000 to 400 $cm^{-1}$, and the lower panel shows the enlarged 4000-400 $cm^{-1}$ region. Spectra are vertically offset for clarity. The broadband FTIR spectra shown here were newly acquired from archived powder samples produced in 2016. The full FTIR dataset is available in the "Data behind Figure" files.

Although the overall spectral profiles remain broadly similar throughout the $CH_4$-series samples, systematic variations in the relative absorption strengths are observed with increasing $CH_4$ abundance. In particular, the C-H stretching features near 3000-2800 $cm^{-1}$ become progressively stronger at higher $CH_4$ concentrations, while the relative prominence of nitrogen-bearing absorptions decreases. The broad absorption complex between 1800 and 1000 $cm^{-1}$ also exhibits moderate changes in shape and relative intensity across the $CH_4$ range, indicating gradual compositional evolution of the haze particles as methane abundance increases. These trends are consistent with the decreasing N/C ratios and increasing hydrocarbon content previously reported for the same samples by C. He, et al. (2022b), as well as with the density variations discussed in **Section 3.1**. Previous laboratory studies have similarly shown that increasing $CH_4$ abundance in $N_2$-dominated plasma systems promotes the formation of more hydrocarbon-rich organic solids with reduced nitrogen incorporation (e.g., C. He, et al. 2022b; S. M. Hörst, et al. 2018). The spectral evolution observed here therefore likely reflects a progressive shift from relatively nitrogen-rich to increasingly carbon-rich haze chemistry at higher methane concentrations. However, because the absolute haze mass incorporated into the pellets varies among samples, the infrared absorbance spectra alone do not provide a strictly quantitative comparison of functional-group abundances. A more quantitative comparison of the wavelength-dependent absorption strength is therefore discussed in **Section 3.3** using the derived imaginary refractive index $k$. Because several broad mid-infrared regions contain overlapping contributions from multiple functional groups, the observed spectral changes are interpreted in terms of variations in the overall absorption envelope rather than as unique quantitative changes in individual functional-group abundances.

The FTIR spectra of solid products generated from $N_2/CH_4/CO$ gas mixtures with fixed $CH_4$ concentration (5%) and varying CO abundances (0-5%) are shown in **Figure 3b**. Overall, the spectra are broadly similar to those of the $CH_4$-series samples, indicating that the major classes of functional groups remain present throughout the experiments. Several broad absorption regions are consistently observed, including a strong N–H/O-H stretching region between 3500 and 2500 cm$^{-1}$, C-H stretching bands near 3000-2800 cm$^{-1}$, absorptions around 2250-2150 cm$^{-1}$ associated with triple-bonded species and cumulated double bonds, and a broad strong double bond absorption between 1800 and 1500 cm$^{-1}$. Additional weaker features are present throughout the fingerprint region below 1500 cm$^{-1}$.

Although the overall spectral shapes remain similar, systematic spectral evolution is observed with increasing CO concentration. The broad absorption features centered near 3500-2500 cm$^{-1}$ and 1800-1500 cm$^{-1}$ become progressively broader as CO abundance increases, consistent with enhanced contributions from oxygen-bearing and polar functional groups. In contrast, the relative intensity of the C-H stretching bands near 3000-2800 cm$^{-1}$ gradually decreases with increasing CO concentration. The strong absorption feature near 2250–2150 cm$^{-1}$ remains prominent across the full CO range, consistent with the continued presence of triple-bonds and cumulated double bonds within the haze materials. Absorptions within the 1500-1000 cm$^{-1}$ region become increasingly pronounced in the CO-series samples, suggesting enhanced contributions from oxygen-containing species and increased complexity of the haze materials. The broad absorption regions contain overlapping contributions from oxygen-, nitrogen-, and hydrocarbon-bearing functional groups. Their spectral evolution is therefore discussed as qualitative evidence for overall compositional changes, together with the previously reported elemental and molecular characterization of the same samples.

The observed spectral evolution is broadly consistent with previous laboratory studies of CO-containing haze analogs (L. Jovanović et al. 2020; Z. Yang et al. 2026; Z. Yang et al. 2025). Elemental analyses on the same set of samples reported by C. He, et al. (2017) showed that oxygen incorporation into the haze particles increases systematically with CO concentration, resulting in higher O/C ratios and progressively more oxygen-rich organic material. Recent high-resolution mass spectrometry (HRMS) analyses further suggested that CO chemistry promotes the formation of chemically complex CHON species with increased oxygen content and molecular unsaturation (Jovanović et al. 2020; Yang et al. 2025). The spectral broadening observed here, particularly in the 1800-1000 cm$^{-1}$ region, is therefore consistent with increasingly oxygen-rich and chemically complex haze compositions at higher CO abundances. Together with the density trends discussed in **Section 3.1**, these results indicate that CO substantially modifies the chemical composition and molecular structure of the resulting haze analogs. As discussed above, variations in pellet thickness and haze loading prevent the FTIR spectra from providing strictly quantitative comparisons of absorption strength among different samples. The spectra presented here are therefore primarily used to identify functional groups and compositional trends. A more quantitative comparison of the wavelength-dependent absorption properties is presented below using the derived imaginary refractive index $k$.

### 3.3 Optical Constants of Haze Analogs

The optical constants ($N = n + ik$) of the haze analogs are derived over 0.4–28.6 μm. Results are presented and discussed over 0.4–25 μm in the main text; the full dataset through 28.6 μm is provided in the Data behind Figures files. This wavelength range significantly extends or complements previous laboratory measurements of haze optical constants (**Table 1**), which are typically limited to narrower intervals (e.g., 0.2-0.9 μm, 0.4-3.5 μm, or ≤15 μm), while partially overlapping with several recent datasets extending to longer wavelengths. The present measurements therefore provide self-consistent optical constants spanning key wavelength regions relevant to radiative transfer calculations and interpretation of observations of planetary atmospheres with similar compositions. **Figures 4** and **5** show the derived optical constants of the archived haze analogs from the $CH_4$- and CO- series production campaigns, respectively, with the left panels presenting real refractive index ($n$) and the right panels presenting the imaginary refractive index ($k$).

**Figure 4a** shows that the retrieved $n$ values of the haze analogs from the $CH_4$-series generally vary smoothly across most wavelengths, with values typically ranging from ∼1.2 to 1.7 under the common anchor reference. Notable decreases in the $n$ values occur near strong absorption regions in the $k$ spectra, particularly around 5–6.5 μm. Across samples, the $n$ values exhibit slightly larger differences between 7 and 12 μm, which is therefore enlarged in **Figure 4c**. The largest separation of n between samples is about 0.05. We do not interpret this modest inter-sample variation in $n$ as quantitative composition-dependent differences in absolute refractive index, because all retrieved $n$ spectra are referenced to the same assumed SKK anchor value with associated anchor uncertainty of 0.069. In addition, if the true refractive index at the anchor frequency differs among compositions, the retrieved $n$ spectra would differ by a wavelength-independent offset. Measurements of related $N_2/CH_4$ tholins by A. Mahjoub, et al. (2012) show $n$ values of approximately 1.48-1.58 near 0.63 μm for samples produced with 1-10% $CH_4$. Differences in production conditions and optical methods preclude a direct quantitative transfer to the present samples; nevertheless, this range illustrates that composition-dependent anchor offsets could potentially be on the order of 0.05-0.10. We therefore focus on the broader wavelength-dependent behavior across the $CH_4$ series rather than on small absolute separations between individual curves. Small differences among neighboring n spectra should be interpreted in the context of the anchor-frequency sensitivity range shown in **Figure 4a.**

As shown in **Figure 4b,** the $k$ values for the $CH_4$-series samples vary by more than two orders of magnitude across the measured wavelength range, from below 0.003 in weakly absorbing regions to peak values of ~0.25-0.28 in the strongest infrared absorption bands. The spectral shape of derived $k$ remains broadly similar among samples and reflects the FTIR absorption features discussed in **Section 3.2**. Three major absorption regions are enlarged in **Figure 4d-f**, corresponding to the FTIR absorption envelopes assigned to C-H/N-H/O-H stretching, triple-bonded/cumulated double-bonded groups, and double-bonded groups, respectively. These assignments identify the dominant contributors to each spectral region; however, the overlapping bands do not permit a unique quantitative separation of individual functional groups.

Despite the overall similarity in trends and absorption features, systematic variations in the relative absorption strengths are observed. The derived $k$ spectra accounted for the haze mass variations in the pellets used for FTIR measurements, enabling a more quantitative comparison of the wavelength-dependent absorption strengths. As $CH_4$ increases, the C-H stretching feature near 3.4 μm becomes progressively more pronounced, consistent with the enhanced hydrocarbon content inferred from the FTIR spectra in **Section 3.2** and previous elemental analyses reported by C. He, et al. (2022b). Consistent with the FTIR spectra, nitrogen-bearing absorption features make comparatively weaker relative contributions as $CH_4$ abundance increases. The absorption feature near 4.5-4.7 μm, associated with triple-bonded nitrogen-bearing functional groups such as nitrile-related species, exhibits a general increase in absorption strength with increasing $CH_4$ abundance. The broad absorption feature near 6-7 μm also varies moderately among the samples. These non-linear trends are consistent with the decreasing particle densities discussed in **Section 3.1**, where higher $CH_4$ concentrations likely promote the formation of more hydrocarbon-rich and less tightly bonded organic solids.

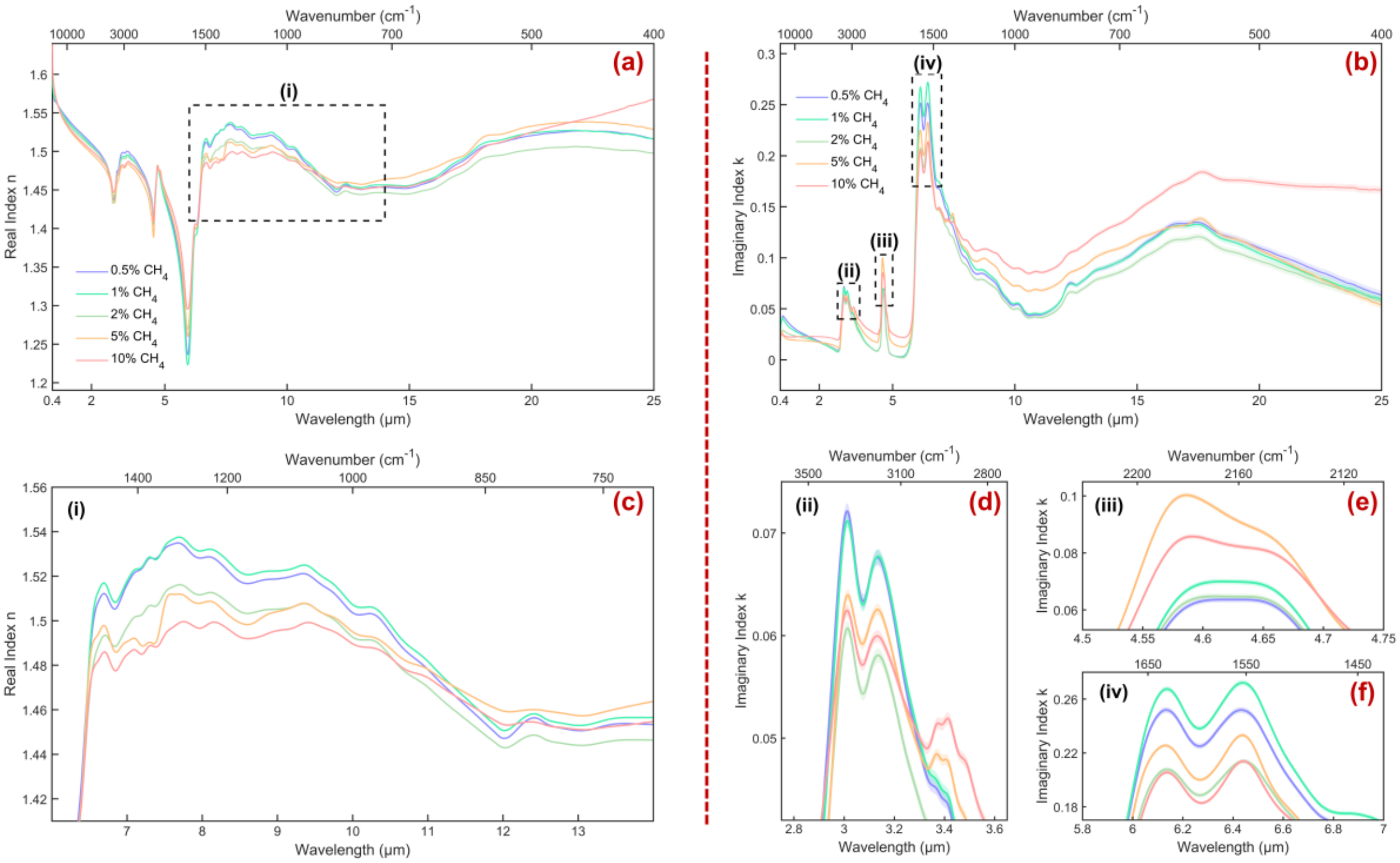

**Figure 4.** Optical constants of $N_2/CH_4$ haze analogs with $CH_4$ concentrations of 0.5-10% over the 400-25000 cm$^{-1}$ (0.4-25 μm) spectral range: Panels (a) and (b) show $n$ and $k$, respectively; panels (c) and (d–f) show enlarged spectral regions. Shaded bands in $k$ denote propagated ±1σ uncertainties. The full optical-constant dataset (0.4-28.6 μm) with associated uncertainties is available in the "Data behind Figure" files. The anchor uncertainty is common to all retrieved $n$ spectra rather than being sample-specific. Therefore, it is not shown as a separate uncertainty band in panels (a) and (c), but is provided in the "Data behind Figure" files.

For the CO-series samples, **Figure 5a** shows that the $n$ values exhibit a smooth trend of variation with different CO mixing ratios in the initial gas mixture, ranging from 1.2 to 1.7. Similar to the $CH_4$-series sample, the decrease in $n$ occurs near the strong absorption regions in the corresponding $k$ spectra. The differences among samples become more apparent in the 7-12 μm region, but remain small, with a maximum difference of only ~0.03 (**Figure 5c**). Similar to the $CH_4$-series samples, the retrieved $n$ spectra for the CO-series sample are also subject to the common-anchor assumption. Given the small difference among CO-series sample, these variations are insufficient to establish a clear trend with the initial gas composition, because the true refractive indices at the anchor frequency may differ among samples and the retrieved values inherit the uncertainty associated with the independently determined anchor value.

The $k$ values exhibit more substantial spectral evolution with increasing CO abundance (**Figure 5b**). The broad absorption region centered near 3 μm becomes progressively wider, consistent with enhanced contributions from hydrogen-bonded N-H and O-H functional groups in increasingly oxygen-rich materials. In contrast, the absorption feature near 4.5-4.7 μm generally decreases in relative strength with increasing CO abundance, suggesting that oxygen incorporation modifies the balance of nitrogen-bearing unsaturated chemistry and reduces the relative contribution of triple-bonded functional groups. The strongest compositional dependence occurs in the 6–7 μm region, where the $k$ spectra become both broader and more intense as CO concentration increases. This broad region contains overlapping contributions from carbonyl-, nitrogen-, and unsaturated functional groups. Its evolution is therefore interpreted as a change in the combined absorption envelope, rather than as a unique change in any individual component. Similarly, the absorption region between 8 and 10 μm becomes increasingly pronounced in the CO-rich samples, consistent with stronger contributions from C-O stretching vibrations identified in the FTIR spectra. The corresponding $k$ spectra, derived from the FTIR transmittance measurements, show the same overall evolution: increasing CO abundance is associated with broader oxygen-bearing absorption envelopes and weaker relative contributions from C-H and triple-bonded features. Together with the previously reported elemental analyses, the spectral trends are consistent with altered nitrogen incorporation, including reduced contributions from unsaturated nitrogen-bearing groups (C. He, et al. 2017). However, these trends are not strictly linear with CO mixing ratio in initial gases. In particular, the 1% CO sample also differs from the otherwise gradual spectral evolution in several regions. This behavior may reflect non-linear compositional or structural changes in the haze material; however, we do not interpret it as evidence for a sharp compositional threshold.

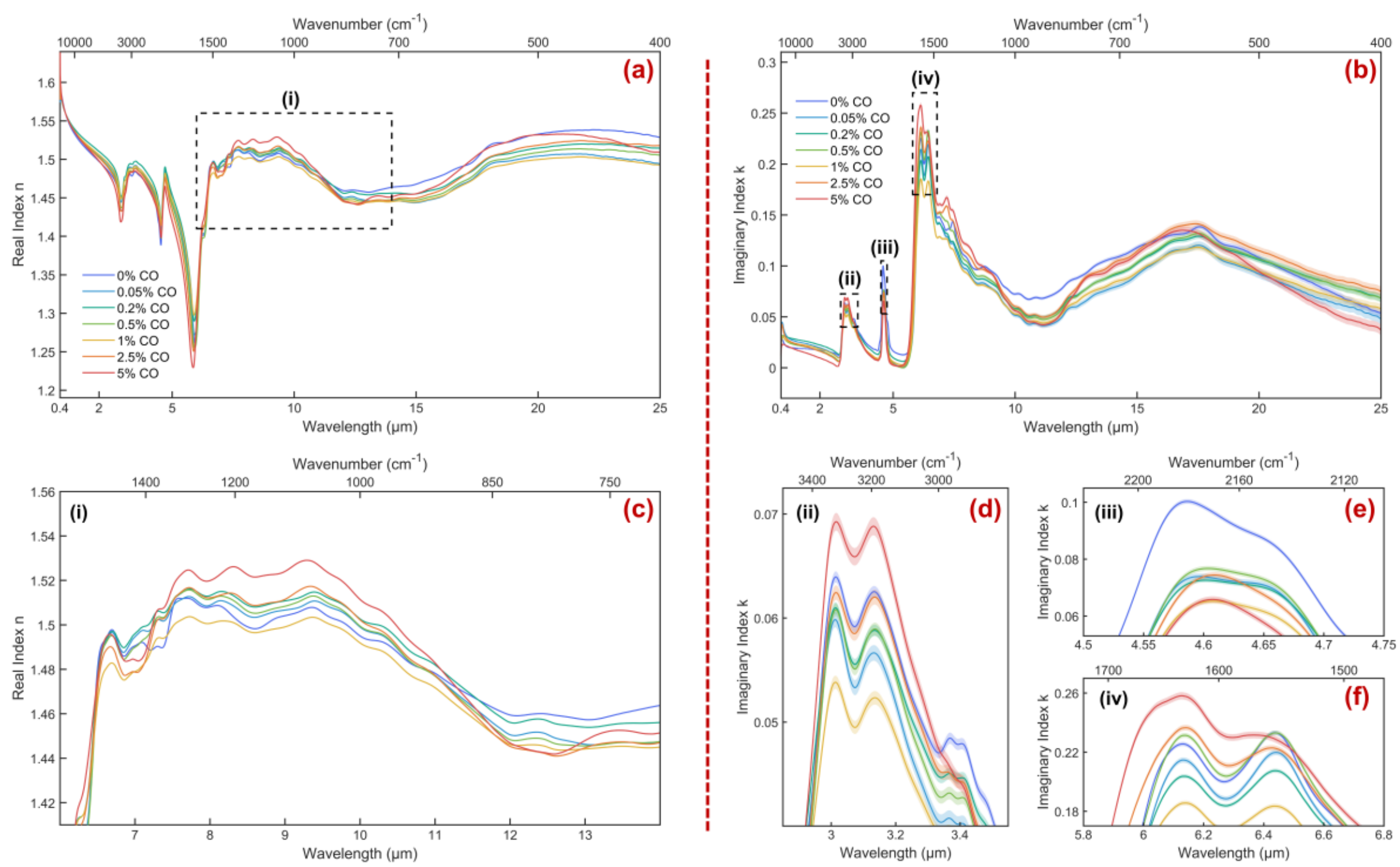

**Figure 5.** Optical constants of $N_2/CH_4/CO$ haze analog gas mixtures with fixed $CH_4$ concentration (5%) and varying CO abundances of 0%, 0.05%, 0.2%, 0.5%, 1%, 2.5%, and 5% over the 400-25000 $cm^{-1}$ (0.4-25 μm) spectral range: Panels (a) and (b) show $n$ and $k$, respectively; panels (c) and (d–f) show enlarged spectral regions. Shaded bands in $k$ denote propagated ±1σ uncertainties. The full optical-constant dataset (0.4-28.6 μm) with associated uncertainties is available in the "Data behind Figure" files. The anchor uncertainty is common to all retrieved $n$ spectra rather than being sample-specific. Therefore, it is not shown as a separate uncertainty band in panels (a) and (c), but is provided in the "Data behind Figure" files.

Our results can be placed in the context of previous optical-constant studies of laboratory haze analogs. A. Mahjoub, et al. (2012) reported that for $N_2/CH_4$ tholin films, increasing the initial $CH_4$ abundance from 1% to 10% was accompanied by an increase in $n$ near 1 μm from approximately 1.48 to 1.58 and a roughly one-order-of-magnitude decrease in $k$ in the same range. For CO-containing Pluto haze analogs, L. Jovanović, et al. (2021) found that increasing the $CH_4/N_2$ ratio at fixed CO abundance was accompanied by lower $n$ and lower UV–visible $k$. These studies demonstrate that the initial $CH_4$ abundance can modify haze optical properties, although the magnitude and direction of the response may depend on production conditions and spectral range.

The stronger and broader infrared absorption envelopes observed in the CO-rich samples are also qualitatively consistent with the enhanced and broadened infrared absorption reported for more oxygenated $CO_2$-containing haze analogs by T. Drant, et al. (2024). This comparison is qualitative, because $CO_2$ and CO do not necessarily lead to identical chemical pathways. A recent cross-laboratory study also found that variations in the $N_2/CH_4$ ratio can strongly affect haze optical properties, whereas the effect of CO was modest under its specific production conditions (T. Drant, et al. 2026). Direct quantitative comparisons among these studies are therefore not appropriate because of differences in gas compositions, haze production conditions, sample forms, wavelength ranges, and optical retrieval methods. Nevertheless, the collective results support the conclusion that both the carbon-source abundance and the oxygen-bearing chemistry of the initial gas mixture play important roles in determining the optical properties of haze.

In summary, for both the $CH_4$- and CO-series samples, the $n$ values range from ~1.2 to 1.7 with modest composition-dependent variations, while $k$ values exhibit strong wavelength-dependent absorption features associated with specific functional groups. Increasing $CH_4$ enhances hydrocarbon-related absorption features, whereas increasing CO broadens and strengthens the 6–10 μm region and weakens the 4.6 μm feature. Together, the wavelength-dependent values of $n$ and $k$ influence the scattering and absorption properties of photochemical hazes, thereby affecting atmospheric radiative transfer, temperature structure, photochemistry, and the vertical distribution of clouds and aerosols in hazy planetary atmospheres. In particular, variations in haze $k$ directly modify the continuum opacity and spectral slopes in transmission and emission spectra. Because the optical constants derived here span a broad wavelength range from the visible to the mid-infrared (0.4–28.6 μm), overlapping with observational wavelength coverage accessible to spectrometers on spacecraft and telescopes such as JWST/NIRSpec (0.6–5.3 μm), JWST/MIRI (5–28 μm), compositionally appropriate haze optical constants are important for accurately interpreting planetary and exoplanetary spectra. The systematic dataset presented here enables, for the first time, composition-consistent haze optical constants for $N_2$-dominated atmospheres with varying $CH_4$ and CO abundances, thereby providing a critical laboratory foundation for accurate radiative transfer modeling and spectral interpretation of hazy Solar System bodies and exoplanets.

Comparison of the controlled $CH_4$- and CO-series provides context for the recent cross-laboratory analysis of T. Drant, et al. (2026). Within the composition ranges investigated here, changes in the initial $CH_4$ abundance produce the more systematic variations in the retrieved optical constants, whereas the CO-series shows smaller but measurable wavelength-dependent differences. This behavior is broadly consistent with the conclusion of T. Drant, et al. (2026) that $N_2/CH_4$ variations exert a stronger influence on haze refractive indices than $CH_4$/CO variations. L. Jovanović, et al. (2021) likewise reported composition-dependent $n$ and $k$ values for CO-containing Pluto haze analogs as the $CH_4/N_2$ ratio was varied. However, differences in production conditions, sample form, wavelength coverage, and optical retrieval method preclude a direct quantitative attribution of the differences between their results and the present controlled series to $CH_4$- or CO-related effects alone. Nevertheless, the present CO-series demonstrates that CO-related chemistry can still modify both $n$ and $k$, particularly in specific infrared spectral regions. Our results therefore support a stronger overall influence of $CH_4$ while showing that CO should not be treated as optically negligible.

### 3.4 Benchmark Comparison Among Titan-, Pluto-, and Triton-like Hazes

The controlled $CH_4$- and CO-series discussed above provide the basis for interpreting the composition-dependent optical behavior of representative planetary haze analogs. We therefore compare Titan-, Pluto-, and Triton-like haze analogs to examine which differences are consistent with the $CH_4$- and CO-dependent trends identified in **Sections 3.2 and 3.3.** The Titan-like analog corresponds to the 5% $CH_4$/95% $N_2$ experiment from previous studies by (C. He, et al. 2022b), while the Pluto-like analog is represented by the 0.05% CO/5% $CH_4$/95% $N_2$ experiment, approximately consistent with the ~500 ppm CO abundance observed in Pluto's atmosphere (C. He, et al. 2017). These samples were produced in 2016; their broadband FTIR spectra were newly measured in this study and their optical constants were derived in the present work. The Triton-like haze is the identical sample reported in previous study (S. E. Moran, et al. 2022), produced using gas mixture consisting of 0.5% CO/0.2% $CH_4$/99.3% $N_2$. For consistency, the optical constants of Triton-like haze shown here were derived using the same procedures adopted in this work. **Figure 6** compares the derived optical constants for these three representative $N_2$- dominated haze analogs. The comparison below uses the $CH_4$- and CO-dependent trends established in the controlled series to interpret the observed benchmark differences.

The real refractive index $n$ of the three benchmark haze analogs show broadly similar wavelength-dependent dispersion, with most values remaining within the range of ~1.2-1.7 throughout the infrared region. Because all $n$ spectra are retrieved using the same SKK reference condition, these differences are interpreted primarily in terms of relative spectral dispersion rather than as independent absolute refractive-index offsets among the three compositions. The $k$ values show more substantial benchmark differences. The Triton-like haze exhibits broader and generally stronger mid-infrared absorption, particularly in the 6–10 μm region. This behavior is broadly consistent with the controlled CO-series result that increasing CO abundance is associated with broader and stronger oxygen-influenced mid-infrared absorption envelopes. The Triton-like sample contains substantially more CO than the Titan- and Pluto-like analogs, and its broad infrared absorption is therefore consistent with a stronger influence of oxygen-bearing haze chemistry.

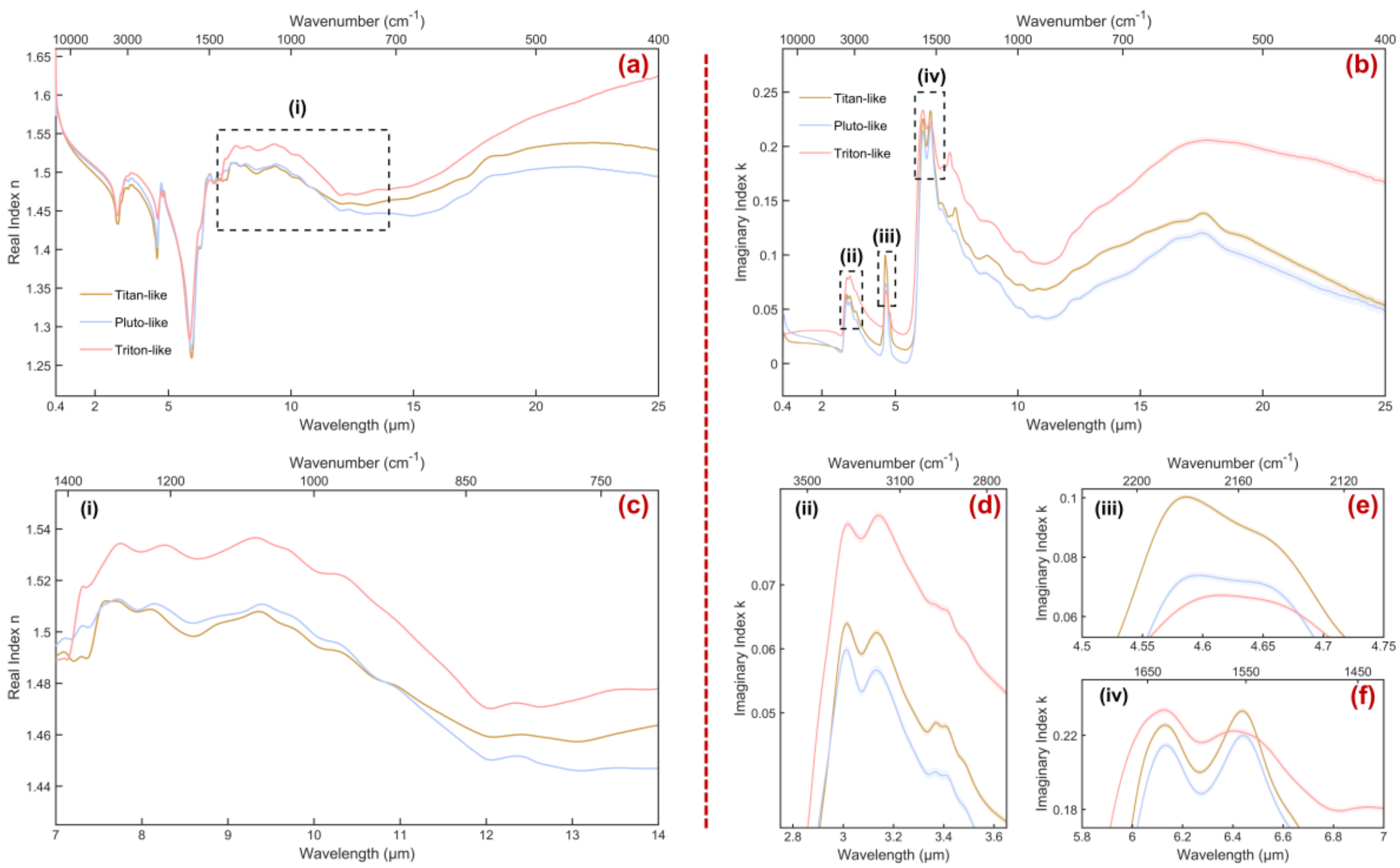

**Figure 6.** Benchmark comparison of the derived optical constants for Titan- , Pluto- , and Triton-like haze analogs over the 400-25000 cm$^{-1}$ (0.4-25 μm). Panels (a) and (b) show $n$ and $k$, respectively; panels (c) and (d–f) show enlarged spectral regions. The Titan-like, Pluto-like, and Triton-like initial gas mixtures were 5% $CH_4$/95% $N_2$, 0.05% CO/5% $CH_4$/95% $N_2$, and 0.5% CO/0.2% $CH_4$/99.3% $N_2$, respectively. The Triton-like sample was produced in 2022 (S. E. Moran, et al. 2022); its broadband FTIR spectrum was newly measured and its optical constants were derived in the present study. Shaded bands in $k$ denote propagated ±1σ uncertainties. The full optical-constant dataset (0.4-28.6 μm) with associated uncertainties is available in the "Data behind Figure" files. The anchor uncertainty is common to all retrieved $n$ spectra rather than being sample-specific. Therefore, it is not shown as a separate uncertainty band in panels (a) and (c), but is provided in the "Data behind Figure" files.

However, the benchmark samples differ in more than one compositional parameter. The Triton-like haze has both lower $CH_4$ and higher CO abundances than the Titan- and Pluto-like analogs; its optical properties therefore cannot be attributed uniquely to CO. The Titan- and Pluto-like analogs have the same initial $CH_4$ abundance but differ by the addition of only 0.05% CO, and their optical differences are correspondingly modest compared with those involving the Triton-like sample. Thus, the benchmark comparison does not independently establish a single-parameter trend. Instead, it illustrates how the $CH_4$- and CO-dependent behavior identified in the controlled experiments can combine in representative $N_2$-dominated planetary haze compositions.

Taken together, the controlled $CH_4$- and CO-series establish that haze optical properties depend on the initial gas composition, while the benchmark comparison illustrates how these trends can help interpret differences among representative Titan-, Pluto-, and Triton-like haze analogs. These results reinforce that a single universal optical-constant dataset is unlikely to represent all $N_2$-dominated haze environments adequately. Compositionally appropriate optical constants are therefore important for atmospheric radiative-transfer calculations and for interpreting spacecraft and telescope observations of hazy planetary atmospheres.

## 4. Conclusions

In this work, we newly measure the broadband vacuum FTIR spectra of the haze analogs and particle densities for the $CH_4$-series and Triton-like samples, and derive the optical constants of previously produced photochemical haze analogs from $N_2/CH_4/CO$ gas mixtures under laboratory plasma conditions. Previously published CO-series particle densities from C. He, et al. (2017) were used as retrieval inputs. The FTIR spectra reveal chemically complex organic solids containing hydrocarbon-, nitrogen-, and oxygen-bearing functional groups. Increasing $CH_4$ abundance enhances aliphatic hydrocarbon features and corresponds to decreasing particle density, whereas increasing CO abundance promotes oxygen incorporation, broader mid-infrared absorptions, and higher particle density. Under the common anchor reference, the retrieved real refractive index $n$ generally remains within ∼1.2-1.7 but shows dispersive variations near strong absorption regions. The derived $k$ spectra exhibit major absorption features near ∼3 μm, ∼4.6 μm, and ∼6-10 μm, associated with N-H/O-H, aliphatic C-H, nitrile-related, carbonyl-bearing, and other nitrogen- and oxygen-containing functional groups. The optical constants show composition-dependent spectral variations, with the most robust differences occurring in the major infrared absorption regions. Smaller differences between neighboring compositions should be interpreted in the context of the propagated uncertainties and the common-anchor sensitivity of the SKK retrieval.

The controlled $CH_4$- and CO-series establish the composition-dependent optical behavior of the haze analogs. The benchmark comparison then shows that the narrower hydrocarbon-related absorption of the Titan-like haze and the broader, stronger mid-infrared absorption of the Triton-like haze are broadly consistent with these controlled trends. Because the benchmark compositions differ in both $CH_4$ and CO abundance, the comparison is used as an interpretation of the combined compositional effects rather than as an independent test of a single parameter.

The optical constant datasets presented here provide improved laboratory constraints for atmospheric radiative transfer models and spectroscopic retrieval studies involving photochemical hazes. Because the strongest haze absorption regions overlap substantially with wavelength ranges accessible to spacecraft and telescopes like HST and JWST, compositionally appropriate haze optical constants will be important for interpreting spectra of hazy planetary and exoplanetary atmospheres.

## Acknowledgments

The authors gratefully acknowledge the support from the National Key Research and Development Program of China (2025YFF0811800) and National Natural Science Foundation of China (42475132).